\documentclass[%
 reprint,
 amsmath,amssymb,
 aps,
]{revtex4-2}

\usepackage{graphicx}% Include figure files
\usepackage{dcolumn}% Align table columns on decimal point
\usepackage{bm}% bold math
\usepackage{appendix}
\begin{document}

\preprint{APS/123-QED}

\title{Finite Measured Entropy Production \\ Does Not Imply Finite Microscopic Irreversibility}% Force line breaks with \\

\author{Hieu Nguyen}
 \email{hieutn@bu.edu}
\affiliation{%
 Faculty of Computing and Data Sciences, Boston University \\
 Boston, Massachusetts 02215, USA
}%
\date{\today}% It is always \today, today,

\begin{abstract}
Coarse-grained measurements provide a lower bound on entropy production, but how much microscopic irreversibility can they conceal? We construct a driven quantum system in which an imperfect counter registers a finite, strictly positive arrow of time while the microscopic entropy production is infinite. The system is a degenerate parametric oscillator below threshold, coupled to a thermal reservoir through repeated interactions with bosonic ancillas. The oscillator sustains a finite, positive heat current into the vacuum bath, where absorption is forbidden and forward and reverse trajectory laws have different supports. A photon counter with finite collection efficiency, gated time bins, and independent dark counts defines a common coarse-graining of the aligned trajectory laws. For fixed observation duration $\mathcal T$ and detector parameters, we prove that the observed relative entropy remains uniformly bounded as the bath temperature $T_b$ vanishes and is strictly positive at vacuum. The hidden contribution obeys \(\Sigma_{\mathrm{hid}}(\mathcal T,T_b)=\mathcal T\dot Q_0/(k_BT_b)+O(1)\), where \(\dot Q_0>0\) is the stationary vacuum heat current. False positives conceal the support mismatch while preserving measurable time asymmetry. Thus, finite and informative measurements of entropy production can coexist with infinite microscopic irreversibility.
\end{abstract}

\maketitle
%\tableofcontents
\section{\label{sec:level1}Introduction\protect }
In thermodynamics, it is the production of entropy that gives the direction of the arrow of time. To be more precise, if the probability of a trajectory moving forwards in time is different than the probability of the time reversed trajectory, then the system has produced entropy \cite{Seifert2005, Kawai_2007,Parrondo_2009,Landi_2021}. Following the second law, the mean entropy production is nonnegative, and it is this average that lets an observer distinguish forward time from backward time; individual trajectories may still have a negative log ratio. Irreversibility is thereby a statement about distinguishability, namely how reliably an observer who sees a trajectory can tell whether it was played forward or backward. In experiments on small systems, these histories are accessible through records produced by particular measurement devices. However, detectors are imperfect and can miss signal events, sometimes even register false positives. Every record an experimenter keeps is a coarse-graining of the microscopic history. The data-processing inequality guarantees that the relative entropy of the coarse records cannot exceed that of the fine ones, so a measured entropy production is a lower bound on the microscopic one. The problem is to determine how much of this microscopic asymmetry survives observation, and how much can remain hidden.

The microscopic details of the system can be made more precise using open quantum systems. In quantum-jump thermodynamics, each jump transfers a definite energy between system and reservoir. Local detailed balance assigns it a reservoir entropy fixed by the ratio of forward and backward jump rates \cite{Landi_2021,Horowitz_2012,Horowitz_2013,Manzano_2018}. What this means is that the system is allowed to emit quanta of energy $\hbar\omega_c$ into a bath at temperature $T_b$ while also absorbing quanta from it. The emission contributes $\hbar\omega_c/k_BT_b$ to the forward–reverse log-likelihood ratio, and this contribution grows without bound as the bath cools. At zero temperature, the system can no longer absorb energy from the bath because the bath is in its vacuum state and thus has no excitations to supply. The consequence of this is that the forward jump is allowed to have non-zero probability while the backward jump is strictly forbidden. The specific problem we address is therefore the following: given a physically specified, imperfect counter observing a driven dissipative quantum system, how much of the system’s irreversibility does the counter register? How much remains hidden, and how does the hidden part grow as the bath approaches its vacuum state?

The part of the entropy production that a coarse record fails to register is called hidden entropy production, and its history in classical stochastic thermodynamics has been well documented \cite{Seifert2005,Kawaguchi_2013,Esposito_2012,mehl2012rolehiddenslowdegrees}. Divergent entropy production also appears classically when a transition has no reverse \cite{Zeraati_2012,Rahav_2014,Busiello_2020}. However, in each of these approaches, the reverse dynamics or the entropy observable is modified so that the result stays finite. In quantum systems the corresponding singularities are characterized by absolute irreversibility, which occurs when forward and reversed path measures have different supports. This means that the forward path could have positive probability while the reversed path has zero probability. On such paths, we say that the relative entropy is infinite, as observing an event with zero probability would produce infinite surprise. Murashita, Funo and Ueda derived the fluctuation equalities that hold under absolute irreversibility \cite{Murashita_2014,Funo_2015}, Manikandan, Elouard and Jordan showed that continuous measurement creates such paths through wave-function collapse \cite{Manikandan_2019}, Yuge and Yamaguchi removed the divergent contribution of a zero-temperature drain in cavity quantum electrodynamics \cite{Yuge_2020}, and Kewming and Shrapnel replaced the divergent inverse temperature of a homodyne-monitored cavity at zero temperature by an effective one \cite{Kewming_2022}. Recent work by Green, Gupta, and Premkumar has provided an instance of this infinite entropy production. For their phase space description of stochastic inflation, the path entropy production is defined as the relative entropy between forward and momentum-reversed paths. There, momentum friction without momentum noise makes the total entropy production infinite\cite{green2026quantummechanicsrareevents}. In a worked example, a walker on a ring driven by a constant force, they find that a position-basis description retains the steady-state entropy production while a momentum-basis description shows none. They conclude that projecting onto a measurement basis can hide steady-state entropy production. What is missing from their example is an operational framework, the closest of which could be provided by Ferri-Cortés et al's model of imperfect detection of quantum jumps through channel efficiencies \cite{Ferri_Cort_s_2025}. The limitation is that their detector has efficiencies only, and a detector characterized only by efficiencies also cannot conceal the vacuum divergence. That is, a detected emission still has zero reverse probability, so the visible relative entropy remains infinite.

Neither work provides a physically realizable model whose microscopic entropy production is infinite, an explicit detector that conceals the divergence while still registering a positive arrow of time, nor a quantitative law for the concealed part. In this paper, we will construct such a model and prove such a law. The system is a single mode of a degenerate parametric oscillator driven below threshold and damped by a thermal bath at temperature $T_b$ \cite{Walls_2008,Drummond1980}. We realize the bath as a sequence of collisions with thermal ancillas. The fully resolved record is then the sequence of ancilla outcomes, and any detector is a classical post-processing of it. The stationary state carries a finite, positive heat current $\dot Q_0$ into the bath even at $T_b=0$. At vacuum the oscillator therefore dissipates at a finite rate while, as we show, its path entropy production is infinite. The central result is the low-temperature law for the hidden part. For a photon counter with finite collection efficiency, gated time bins and dark counts, we prove that the observed relative entropy stays finite and strictly positive at vacuum, while the hidden part grows as $\Sigma_{\rm hid}(\mathcal T,T_b)=\mathcal T\dot Q_0/k_BT_b+O(1)$ as $T_b\to0$. Missed detections and coarse time resolution cannot conceal the divergence; only false positives can, and they do so while the counter still registers an arrow of time.

\section{Physical Model and Trajectory Thermodynamics}
In this section, we provide the model we will be analyzing and all of the assumptions we need. We define the driven oscillator along with its bath, and realize the bath as a stream of collision ancillas. This will allow us to derive an exact trajectory identity that connects the forward–reverse log-likelihood ratio to the heat delivered to the bath. Each assumption is stated where it enters. 
\subsection{The Driven Oscillator}
We consider a single bosonic cavity mode with annihilation operator \(a\), satisfying \([a,a^\dagger]=1\), and carrier frequency \(\omega_c>0\). A classical pump at frequency \(\omega_p\approx2\omega_c\) creates and annihilates cavity excitations in pairs. We assume that the pump is undepleted, so its amplitude is known and its backreaction can be neglected. The laboratory-frame Hamiltonian is
\begin{equation}
\label{labHamiltonian}
    H_{\rm lab}(t)=\hbar\omega_c a^\dagger a+\frac{i\hbar g}{2}\Bigl[e^{i(\varphi-\omega_pt)}a^{\dagger2}-e^{-i(\varphi-\omega_pt)}a^{2}\Bigr]
\end{equation}
where $g\ge0$ is the pump strength, proportional to the pump amplitude and the nonlinear susceptibility, and $\varphi$ is the pump phase. Without loss of generality, we can set the phase to be $\varphi=0$. Let \(R(t)=\exp[i(\omega_p/2)t\,a^\dagger a]\), and define the rotating-frame state by \(\rho_{\mathrm{rot}}=R\rho_{\mathrm{lab}}R^\dagger\). The transformed Hamiltonian is
\begin{align}
H_\varphi
&=
R H_{\mathrm{lab}}R^\dagger+i\hbar\dot R R^\dagger \notag \\
&=
\hbar\Delta a^\dagger a
+
\frac{i\hbar g}{2}(a^{\dagger2}-a^2),
\qquad
\Delta=\omega_c-\frac{\omega_p}{2}.
\end{align}
This allows us to cancel the explicit time dependence. In terms of the quadratures $Q=(a+a^\dagger)/\sqrt2$ and $P=(a-a^\dagger)/(i\sqrt2)$,
this becomes
\begin{equation}
    \frac{H_\varphi}{\hbar}
=
\Delta a^\dagger a+\frac g2\{Q,P\},
\end{equation}
\begin{figure*}
\includegraphics{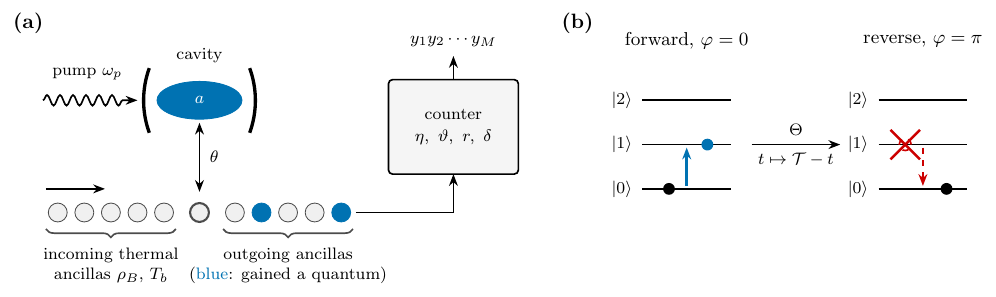}% Here is how to import EPS art
\caption{\label{fig:wide}Setup and the origin of the vacuum singularity. (a) A degenerate parametric oscillator, pumped at frequency $\omega_p$, exchanges quanta with a stream of harmonic ancillas through an exchange coupling of strength $\theta=\sqrt{\gamma\tau_c}$ per collision. Incoming ancillas are prepared in the thermal state $\rho_B$ at temperature $T_b$. Outgoing ancillas that gained a quantum, which are emissions into the bath, are shaded blue. The outgoing ancillas are read by a counter that reports one bit $y_j$ per bin of width $\delta$, with collection efficiency $\eta$, live-bin probability $\vartheta$ and dark-count probability $r$. (b) Ancilla occupations in a forward emission and in its time reverse. In the forward experiment an ancilla arriving in $|0\rangle$ leaves in $|1\rangle$. Time reversal $\Theta$ with $t\mapsto\mathcal T-t$ maps this event to a reverse experiment in which an ancilla must arrive in $|1\rangle$ and leave in $|0\rangle$. At $T_b=0$ every ancilla arrives in $|0\rangle$, so the reverse transition has zero probability.}
\end{figure*}
so the pump squeezes the field along the diagonal quadratures while the detuning rotates it. We assume quasi-resonance, $|\Delta|,g,\gamma\ll\omega_c$. The rotating-wave approximation then holds, and each quantum exchanged with the bath carries energy $\hbar\omega_c$ up to relative corrections of order $|\Delta|/\omega_c$. The total damping rate is \(\gamma>0\). When external and internal loss channels couple proportionally to \(a\) and have the same thermal occupation, their rates add as $\gamma=\gamma_{\mathrm{ext}}+\gamma_{\mathrm{int}}.$ With the thermal damping that we will derive in the next section, the mean field obeys
\begin{equation}
\label{meanfield}
    \frac{d}{dt}\begin{bmatrix}\langle a\rangle\\ \langle a^\dagger\rangle\end{bmatrix}
=\begin{bmatrix}-\gamma/2-i\Delta & g\\ g & -\gamma/2+i\Delta\end{bmatrix}
\begin{bmatrix}\langle a\rangle\\ \langle a^\dagger\rangle\end{bmatrix},
\end{equation}
with eigenvalues $-\gamma/2\pm\sqrt{g^2-\Delta^2}$. For $g^2\le\Delta^2$ the square root is imaginary and both eigenvalues have real part $-\gamma/2$. For $g^2>\Delta^2$ stability requires $\sqrt{g^2-\Delta^2}<\gamma/2$. The two cases combine into the single condition
\begin{equation}
\label{oscillatorThreshold}
    \mathfrak d\equiv\gamma^2+4\Delta^2-4g^2>0.
\end{equation}
The boundary $\mathfrak d=0$ is the oscillation threshold, which at zero detuning is $g=\gamma/2$. Above it, the quadratic model has no stationary state and pump depletion must be included. The drift does not depend on the bath temperature, so \eqref{oscillatorThreshold} is the same at every $T_b$. We assume $\mathfrak d>0$ and $g>0$ throughout, which implies that our states will be Gaussian. 
\subsection{The Thermal Bath}
We realize the reservoir as a sequence of independent bosonic ancillas \(b_n\), each used once (as shown in Fig.~1). Their bare Hamiltonian is
\(
H_{b_n}=\hbar\omega_c b_n^\dagger b_n.
\) This fixes the energy assigned to a bath quantum at the carrier energy \(\hbar\omega_c\), including in the rotating-frame description. For \(T_b>0\), define
\begin{equation}
    \epsilon=e^{-\hbar\omega_c/(k_BT_b)},
\quad
\bar n_\epsilon=\frac{\epsilon}{1-\epsilon}.
\end{equation}
Each ancilla is prepared in the Gibbs state
\begin{equation}
    \rho_B=(1-\epsilon)\sum_{k\ge0}\epsilon^k|k\rangle\langle k|
\end{equation}
During the interval $[(n-1)\tau_c,n\tau_c)$, the oscillator interacts with the single ancilla $n$. The collision unitary is then
\begin{equation}
\label{unitary}
    U_\varphi=\exp\Bigl[-\frac{i\tau_c}{\hbar}\bigl(H_\varphi+\hbar\Delta\,b_n^\dagger b_n\bigr)-i\theta\bigl(ab_n^\dagger+a^\dagger b_n\bigr)\Bigr].
\end{equation}
with $\theta=\sqrt{\gamma\tau_c}.$ The exchange coupling conserves $a^\dagger a+b_n^\dagger b_n$, so the ancilla gains or loses the cavity’s quanta one at a time. The ancillas can be identified with successive time bins of duration $\tau_c$ of the traveling field that couples to the cavity at rate $\gamma$ \cite{Ciccarello_2022,Strasberg_2017}. Physically, this means that the occupation of an outgoing ancilla is the photon number in one time bin of the output field. We use harmonic ancillas so that the coupling $\theta$ does not depend on temperature. 

Now, because $\rho_B$ is diagonal in the number basis, every term of odd order in $b_n$ and $b_n^\dagger$ vanishes on tracing out the ancilla. The first-order term in $\tau_c$ and the second-order term in $\theta$ then give
\begin{equation} \operatorname{tr}_B\bigl[U_\varphi(\rho\otimes\rho_B)U_\varphi^\dagger\bigr]=\rho+\tau_c\,\mathcal L_{\epsilon,\varphi}\rho+O(\tau_c^2),
\end{equation}
with
\begin{align}
\label{MasterEq}
    \mathcal L_{\epsilon,\varphi}\rho&=-\frac{i}{\hbar}[H_\varphi,\rho]+\mathcal D[L_{\rm e}]\rho+\mathcal D[L_{\rm a}]\rho,\\
    L_{\rm e}&=\sqrt{\gamma(\bar n+1)}\,a,
\quad
L_{\rm a}=\sqrt{\gamma\bar n}\,a^\dagger, \notag 
\end{align}
and $\mathcal D[L]\rho=L\rho L^\dagger-\tfrac12\{L^\dagger L,\rho\}$. Consequently, the weights $\bar n+1$ and $\bar n$ are the thermal averages of $\langle b_nb_n^\dagger\rangle$ and $\langle b_n^\dagger b_n\rangle$. As $\tau_c\to0$ at fixed $\gamma$, the composition of $\mathcal T/\tau_c$ collisions converges to the semigroup $e^{\mathcal T\mathcal L_{\epsilon,\varphi}}$. We write $\mathcal L_\epsilon=\mathcal L_{\epsilon,0}$ for the forward generator and $\tilde{\mathcal L}_\epsilon=\mathcal L_{\epsilon,\pi}$ for the reversed one. We can identify the jump operators through the ancilla matrix elements $M_{k'k}=\langle k'|U_\varphi|k\rangle$, which act on the oscillator. To leading order,
\begin{equation}
    M_{k+1,k}=-i\theta\sqrt{k+1}\,a+\ldots,
\quad
M_{k-1,k}=-i\theta\sqrt{k}\,a^\dagger+\ldots,
\end{equation}
while changes of the ancilla occupation by two or more quanta have probability $O(\gamma^2\tau_c^2)$ per collision. Averaging over the initial occupation with weights $\rho_B(k)$ gives $\sum_k\rho_B(k)M_{k+1,k}\rho M_{k+1,k}^\dagger=\tau_cL_{\rm e}\rho L_{\rm e}^\dagger+O(\tau_c^2)$ and $\sum_k\rho_B(k)M_{k-1,k}\rho M_{k-1,k}^\dagger=\tau_cL_{\rm a}\rho L_{\rm a}^\dagger+O(\tau_c^2)$. A collision in which the ancilla gains a quantum is therefore an emission jump $L_{\rm e}$, and one in which it loses a quantum is an absorption jump $L_{\rm a}$. In the laboratory frame the ancilla Hamiltonian is $\hbar\omega_cb_n^\dagger b_n$, and its number basis is the same in both frames. An emission thus delivers $\hbar\omega_c$ to the bath, and an absorption removes $\hbar\omega_c$. The rate ratio
\begin{equation}
    \frac{\gamma(\bar n+1)}{\gamma\bar n}=\frac1\epsilon=e^{\hbar\omega_c/k_BT_b}
\end{equation}
is local detailed balance. Here it is a consequence of the thermal preparation of the ancillas. At vacuum $\epsilon=0$, every ancilla arrives in $|0\rangle$, so $L_{\rm a}=0$ and absorption is impossible, while $L_{\rm e}=\sqrt\gamma\,a$ remains. We take $\gamma$ to be the total energy damping rate of the mode, and assume that every loss channel, including internal loss, couples to a bath at the same temperature $T_b$. All channels then combine into the single pair. The fraction $\gamma_{\rm ext}/\gamma$ of the emission that leaves through the output coupler toward a detector enters only through the collection efficiency. An internal channel at a different temperature would contribute an additional pair of jump operators with its own Boltzmann factor, and would have to be tracked separately.
\subsection{Resolved Records}
The forward experiment starts in the stationary state $\rho_{\rm ss}$ of $\mathcal L_\epsilon$. We will show that for $g>0$, this state has nondegenerate spectrum
\begin{align}
\label{density}
    \rho_{\rm ss}&=\sum_{j\ge0}p_j|\psi_j\rangle\langle\psi_j|, \\
p_j&=(1-\zeta_\epsilon)\zeta_\epsilon^{\,j},
\quad
|\psi_j\rangle=U_{S_\epsilon}|j\rangle, \notag 
\end{align}
where $U_{S_\epsilon}$ is a Gaussian unitary and $0<\zeta_\epsilon<1$ at every temperature, vacuum included. At $t=0$ the oscillator is measured in the basis $\{|\psi_j\rangle\}$, giving outcome $i$ with probability $p_i$. Each ancilla is measured in its number basis before its collision, with outcome $k_n$, and after it with outcome $k_n'$. The first measurement does not change the ancilla because $\rho_B$ is diagonal. The second does not affect the oscillator because the ancilla is discarded. The monitoring is therefore passive: the unconditional oscillator dynamics is given by the Master equation \eqref{MasterEq} whether or not the outcomes are kept. At $t=\mathcal T$ the oscillator is measured in a basis $\{|\phi_f\rangle\}$ fixed below, giving outcome $f$. With $n_{\mathcal T}=\mathcal T/\tau_c$ collisions, the fully resolved record is
\begin{equation}
    \Gamma=\bigl(i;\,k_1,k_1';\,\ldots;\,k_{n_{\mathcal T}},k_{n_{\mathcal T}}';\,f\bigr),
\end{equation}
with probability
\begin{equation}
    P_\epsilon(\Gamma)=p_i\prod_{n=1}^{n_{\mathcal T}}\rho_B(k_n)\,
\Bigl|\langle\phi_f|M_{k'_{n_{\mathcal T}}k_{n_{\mathcal T}}}\cdots M_{k_1'k_1}|\psi_i\rangle\Bigr|^2,
\end{equation}
where $M_{k'k}=\langle k'|U_0|k\rangle$. We write $N_{\rm e}(\Gamma)$ and $N_{\rm a}(\Gamma)$ for the numbers of collisions in which the ancilla gains or loses a quantum, and
\begin{equation}
    N_{\rm net}(\Gamma)=\sum_{n=1}^{n_{\mathcal T}}(k_n'-k_n)
\end{equation}
for the net number of quanta delivered to the bath. Over a window of fixed duration, multi-quantum collisions have total probability $O(\gamma^2\tau_c\mathcal T)$, so $N_{\rm net}=N_{\rm e}-N_{\rm a}$ in the continuum limit.

Now let $\Theta$ denote oscillator time reversal. It satisfies $\Theta a\Theta^{-1}=a$ and $\Theta b_n\Theta^{-1}=b_n$, hence $\Theta Q\Theta^{-1}=Q$ and $\Theta P\Theta^{-1}=-P$. This is the standard time reversal of an oscillator, and it leaves the thermal ancilla states invariant. The time reverse of a forward collision is $\Theta U_0^\dagger\Theta^{-1}$. Since $\Theta$ is antiunitary, $\Theta e^{iX}\Theta^{-1}=e^{-i\Theta X\Theta^{-1}}$. The detuning and exchange terms in \eqref{unitary} are real in the Fock basis, while $\Theta H_0\Theta^{-1}=H_\pi$. Hence we have 
\begin{equation}
\label{reverse}
    \Theta U_0^\dagger\Theta^{-1}=U_\pi .
\end{equation}
The reversed collision is therefore a physical collision with the same detuning, damping, and ancilla temperature all with the pump phase shifted by $\pi$. In the laboratory frame, $\Theta H_{\rm lab}(\mathcal T-t)\Theta^{-1}$ is again of the form \eqref{labHamiltonian} with phase $\pi-\varphi+\omega_p\mathcal T$. The offset $\omega_p\mathcal T$ is the phase the pump reference accumulates over the window, and it disappears in the frame co-rotating with the pump, in which all our statements are made. In that frame, reversal is $\varphi\mapsto\pi-\varphi$.

The reverse experiment uses the same apparatus with pump phase $\pi$. Fresh thermal ancillas interact one per interval through $U_\pi$ and are measured before and after their collisions. The oscillator starts in the stationary state $\tilde\rho_{\rm ss}$ of $\tilde{\mathcal L}_\epsilon$, the state to which the reversed apparatus relaxes. Then let 
\begin{equation}
    V=\exp\left[\frac{i\pi}{2}\left(a^\dagger a+\sum_nb_n^\dagger b_n\right)\right]
\end{equation}
so that $VaV^\dagger=-ia$ and $Vb_nV^\dagger=-ib_n$. Then $V$ leaves the detuning terms, the exchange coupling, $\rho_B$ and the ancilla number measurements invariant, and reverses the sign of the pump term, so
\begin{equation}
\label{symmetry}
VU_0V^\dagger=U_\pi,
\quad
\tilde\rho_{\rm ss}=U_q\rho_{\rm ss}U_q^\dagger,
\quad
U_q=e^{i\pi a^\dagger a/2}.
\end{equation}
The reverse state therefore has the spectrum $\{p_j\}$, with eigenvectors $|\tilde\psi_j\rangle=U_q|\psi_j\rangle$. At its start the reverse experiment measures the oscillator in $\{|\tilde\psi_j\rangle\}$, giving outcome $f$ with probability $\tilde p_f=p_f$, and at its end it measures in $\{\Theta|\psi_i\rangle\}$. The forward final basis is defined by the same rule, $|\phi_f\rangle=\Theta^{-1}|\tilde\psi_f\rangle$, so that each experiment's final measurement is the time reverse of the other's initial one. 
Both experiments use fresh ancillas at the same temperature. Reverse collision $m$ is paired with forward collision $n=n_{\mathcal T}+1-m$. Its pre- and post-collision outcomes correspond to $k_n'$ and $k_n$, because an ancilla that gained a quantum in the forward experiment must lose one in the reverse experiment. The reversed record is
\begin{equation}
    \tilde\Gamma=\bigl(f;\,k'_{n_{\mathcal T}},k_{n_{\mathcal T}};\,\ldots;\,k_1',k_1;\,i\bigr),
\end{equation}
and the aligned reverse law is $Q_\epsilon(\Gamma)=\tilde P_\epsilon(\tilde\Gamma)$, where $\tilde P_\epsilon$ is the record law of the reverse experiment. Since $\Gamma\mapsto\tilde\Gamma$ is a bijection, $Q_\epsilon$ is a probability law on the same record space as $P_\epsilon$. The fully resolved entropy production over the window is
\begin{equation}
    \Sigma_{\rm full}(\mathcal T,T_b)=D_{\rm KL}(P_\epsilon\Vert Q_\epsilon).
\end{equation}
The symmetry \eqref{symmetry} has a useful consequence in that every element of the reverse experiment, including its endpoint bases, is the conjugate of the corresponding element of the forward experiment. That is, since $\Theta U_q=U_q^\dagger\Theta$, the reverse final basis $\Theta|\psi_i\rangle$ equals $U_q|\phi_i\rangle$. Its record law is therefore identical to the forward one, $\tilde P_\epsilon=P_\epsilon$, and
\begin{equation}
\label{recordlaw}
    Q_\epsilon(\Gamma)=P_\epsilon(\tilde\Gamma).
\end{equation}
The arrow of time is thus invisible in the statistics of either experiment taken alone. It appears only when a forward record is compared with its own reversal, in which emissions become absorptions and the order of collisions is inverted. In particular, $\Sigma_{\rm full}$ and every coarse-grained entropy production derived from it can be evaluated from the forward experiment by itself.
\subsection{Trajectory Entropy Production}
The ancilla matrix elements of the reversed collision follow from \eqref{reverse}. Writing $\Theta=\Theta_S\otimes\Theta_B$ and using the fact that the ancilla number basis is real,
\begin{equation}
    \tilde M_{kk'}\equiv\langle k|U_\pi|k'\rangle=\Theta_S\langle k|U_0^\dagger|k'\rangle\Theta_S^{-1}=\Theta_S M_{k'k}^\dagger\Theta_S^{-1}.
\end{equation}
The reverse experiment applies these operators in inverted order to $|\tilde\psi_f\rangle=\Theta|\phi_f\rangle$ and projects onto $\Theta|\psi_i\rangle$. By antiunitarity, $\langle\Theta x|\Theta y\rangle=\langle y|x\rangle$, so
\begin{align}
    \langle\Theta\psi_i|\tilde M_{k_1k_1'}\cdots\tilde M_{k_{n_{\mathcal T}}k'_{n_{\mathcal T}}}&|\Theta\phi_f\rangle
=\notag \\ &\langle\phi_f|M_{k'_{n_{\mathcal T}}k_{n_{\mathcal T}}}\cdots M_{k_1'k_1}|\psi_i\rangle .
\end{align}
The forward and aligned reverse probabilities therefore contain the same squared amplitude and differ only in their preparation weights
\begin{equation}
    Q_\epsilon(\Gamma)=p_f\prod_{n=1}^{n_{\mathcal T}}\rho_B(k_n')\,
\Bigl|\langle\phi_f|M_{k'_{n_{\mathcal T}}k_{n_{\mathcal T}}}\cdots M_{k_1'k_1}|\psi_i\rangle\Bigr|^2 .
\end{equation}
With $\rho_B(k)=(1-\epsilon)\epsilon^k$, the log-likelihood ratio of a resolved record is
\begin{equation}
\label{loglikelihood}
\varsigma_\epsilon(\Gamma)\equiv\ln\frac{P_\epsilon(\Gamma)}{Q_\epsilon(\Gamma)}
=\varsigma_{\rm end}(\Gamma)+N_{\rm net}(\Gamma)\ln\frac1\epsilon
\end{equation}
with $\varsigma_{\rm end}(\Gamma)=\ln\frac{p_i}{p_f}$. Note that the above identity is independent of both the Hamiltonian and the damping rate. So while Hamiltonian and $\gamma$ might change the probability of each record, the ratio is determined entirely by the preparations of the ancillas and of the endpoints.

Let us examine the physical interpretation of each term in \eqref{loglikelihood}. The bath term is a sum of single-collision contributions $\ln[\rho_B(k_n)/\rho_B(k_n')]$, each the change in the ancilla’s surprisal. Since $\ln(1/\epsilon)=\hbar\omega_c/k_BT_b$ and $\hbar\omega_cN_{\rm net}$ is the energy received by the ancillas,
\begin{equation}
    N_{\rm net}\ln\frac1\epsilon=\frac{\mathcal Q}{k_BT_b},
\qquad
\mathcal Q=\hbar\omega_c N_{\rm net},
\end{equation}
which is the Clausius entropy of the heat $\mathcal Q$ delivered to the bath along the record. The endpoint term is the change of the oscillator’s stochastic entropy $-\ln p_j$ between the two reference states. By \eqref{density},
\begin{equation}
\label{endpoint}
    \varsigma_{\rm end}=(f-i)\ln\frac1{\zeta_\epsilon},
\end{equation}
so $\varsigma_{\rm end}$ counts the change in Williamson excitation number weighted by the effective inverse temperature $\ln(1/\zeta_\epsilon)$ of the stationary state. Because $\zeta_\epsilon>0$ for $g>0$, this term is finite on every record at every temperature, including $T_b=0$, and it has finite moments of all orders. Its mean $B_\epsilon=\langle\varsigma_{\rm end}\rangle$ does not vanish in general. The final outcome $f$ is read in the basis $\{\Theta^{-1}|\tilde\psi_f\rangle\}$, which is not the eigenbasis of the forward final state, so its distribution differs from $\{p_f\}$. At positive temperature the two laws are mutually absolutely continuous, and taking the forward mean of \eqref{loglikelihood} gives
\begin{equation}
\label{fullentropy}
    \Sigma_{\rm full}(\mathcal T,T_b)=B_\epsilon+\langle N_{\rm net}\rangle\ln\frac1\epsilon,
\end{equation}
where $\langle N_{\rm net}\rangle=\mathcal T\dot J(\epsilon)$ is fixed by the stationary net emission rate $\dot J(\epsilon)$. At vacuum, $\rho_B(k')=0$ for every $k'\ge1$. The aligned reverse law therefore assigns zero probability to every record that contains an emission, which is consistent with $\varsigma_0=+\infty$ there.
\section{Stationary State}
This section computes the stationary state of the forward dynamics and the photon currents it carries. It establishes two results. First, the oscillator keeps delivering heat to the bath at a finite, positive rate as the bath approaches its vacuum state. Second, the resolved entropy production grows as the Clausius entropy of that heat current while the endpoint contribution stays bounded. We will make the derivation brief, with all the algebraic details in the appendix. 
\subsection{Stationary Covariance}
Because the generator $\mathcal L_\epsilon$ is quadratic and stable, it has a unique stationary state, which is a centered Gaussian state with covariance matrix $X_{jk}=\tfrac12\langle\{R_j,R_k\}\rangle$, here $R=(Q,P)$. The vacuum therefore has covariance $X=\mathbb I/2$. From Sec. II.A, the quadratures obey $\dot R=AR+\text{noise}$, with drift and diffusion matrices
\begin{equation}
    A=\begin{bmatrix} g-\gamma/2 & \Delta\\ -\Delta & -g-\gamma/2\end{bmatrix},
\qquad
D=\gamma\Bigl(\bar n+\tfrac12\Bigr)\mathbb I .
\end{equation}
From $A$, we see that the pump amplifies $Q$ and attenuates $P$ while the bath damps both while adding noise at the thermal level. The stationary covariance solves the Lyapunov equation $AX+XA^{\mathsf T}+D=0$, whose unique solution for $\mathfrak d>0$ is
\begin{equation}
\label{covariance}
    X_{\rm ss}=\frac{2\bar n+1}{2\mathfrak d}
\begin{bmatrix}\mathfrak a+2\gamma g & -4\Delta g\\ -4\Delta g & \mathfrak a-2\gamma g\end{bmatrix},
\quad
\mathfrak a=\gamma^2+4\Delta^2 .
\end{equation}
Since $\langle a^\dagger a\rangle=\tfrac12\operatorname{tr}X-\tfrac12$ and $\mathfrak a-\mathfrak d=4g^2$, the mean occupation is
\begin{equation}
\label{nss}
    n_{\rm ss}=\bar n+(2\bar n+1)\,\frac{2g^2}{\mathfrak d}.
\end{equation}
The first term is the thermal occupation the bath would impose by itself. The second is the population created by down-conversion and amplified by thermal noise. We see that it does not vanish at vacuum, where $n_{\rm ss}=2g^2/\mathfrak d>0$. In that limit the cavity holds photons only because the pump creates pairs out of the vacuum, as in spontaneous parametric down-conversion. The covariance \eqref{covariance} also fixes the spectrum used in Sec. II.C. Its symplectic eigenvalue is
\begin{equation}
\label{nu}
    \nu_\epsilon=\sqrt{\det X_{\rm ss}}=\Bigl(\bar n+\tfrac12\Bigr)\sqrt{\frac{\mathfrak a}{\mathfrak d}} .
\end{equation}
By Williamson’s theorem, $\rho_{\rm ss}=U_{S_\epsilon}\,\tau_{\nu_\epsilon}U_{S_\epsilon}^\dagger$, where $U_{S_\epsilon}$ is a Gaussian unitary and $\tau_\nu$ is a thermal state with mean occupation $\nu-\tfrac12$. The eigenvalues are therefore $p_j=(1-\zeta_\epsilon)\zeta_\epsilon^{\,j}$, with
\begin{equation}
\label{zeta}
    \zeta_\epsilon=\frac{\nu_\epsilon-\tfrac12}{\nu_\epsilon+\tfrac12}.
\end{equation}
For $g>0$ we have $\mathfrak a>\mathfrak d$, so $\nu_\epsilon>\tfrac12$ and $0<\zeta_\epsilon<1$ at every temperature, including $\epsilon=0$. The stationary state stays strictly mixed even when the bath is pure, because the pump and the damping forces it so. 
\subsection{Photon Current}
The stationary rates of emission and absorption jumps are $j_{\rm e}=\operatorname{tr}(L_{\rm e}\rho_{\rm ss}L_{\rm e}^\dagger)=\gamma(\bar n+1)n_{\rm ss}$ and $j_{\rm a}=\operatorname{tr}(L_{\rm a}\rho_{\rm ss}L_{\rm a}^\dagger)=\gamma\bar n(n_{\rm ss}+1)$. Their difference is the net photon current into the bath
\begin{align}
\label{JCurrent}
    \dot J(\epsilon)&=j_{\rm e}-j_{\rm a}\notag \\&=\gamma(n_{\rm ss}-\bar n)\notag \\ &=\dot J_0\,\frac{1+\epsilon}{1-\epsilon},
\quad
\dot J_0=\frac{2\gamma g^2}{\mathfrak d}.
\end{align}
The total jump activity is
\begin{equation}
    j_{\rm e}+j_{\rm a}=\dot J_0\Bigl(\frac{1+\epsilon}{1-\epsilon}\Bigr)^2+\frac{2\gamma\epsilon}{(1-\epsilon)^2}.
\end{equation}
The net current is balanced by the pump. The pump term of $H_0$ changes the occupation at rate $g\langle a^{\dagger2}+a^2\rangle=2g\,\mathrm{Re}\langle a^2\rangle=g(X_{QQ}-X_{PP})$. By \eqref{covariance} this equals $\dot J(\epsilon)$. Each pair created by the pump is therefore eventually delivered to the bath, and the stationary state is a steady flow of quanta from pump to bath. The factor $(1+\epsilon)/(1-\epsilon)=2\bar n+1$ is thermal stimulation, which means that a warmer bath raises the cavity’s noise floor, amplified by the pump.
\subsection{Vacuum Heat Current}
Each net emission delivers energy $\hbar\omega_c$ to the bath (Sec. II.B), so the stationary heat current is $\dot Q(T_b)=\hbar\omega_c\dot J(\epsilon)$. At vacuum,
\begin{equation}
    \dot Q_0=\hbar\omega_c\,\frac{2\gamma g^2}{\gamma^2+4\Delta^2-4g^2}>0
\end{equation}
which holds for every $g>0$ below threshold. There are three immediate and important consequences. First, at vacuum the emission rate $j_{\rm e}=\dot J_0$ is strictly positive while the absorption rate $j_{\rm a}$ vanishes. The forward process therefore produces, at a finite rate, events whose time reverses the vacuum bath cannot produce. This persistent one-way emission is the physical origin of the singularity. Second, the singularity is not accompanied by any divergence in energetic quantities. The heat current, the occupation, and the stationary entropy $-\operatorname{tr}\rho_{\rm ss}\ln\rho_{\rm ss}$ all remain finite and depend smoothly on $\epsilon$ even at $\epsilon=0$. This means that the mechanism that causes divergences at vacuum must be a statement about distinguishability. Lastly, $\dot Q_0$ grows without bound only as $\mathfrak d\to0$, where the parametric approximation fails and pump depletion must be restored. At fixed $\mathfrak d>0$, the heat current is a finite coefficient. For example, with $\Delta=\gamma$ and $g=0.6\gamma$ we have $\dot J_0\approx0.202\,\gamma$, so the cavity emits about one photon per five damping times. At $g=0$ the current vanishes and the oscillator relaxes to equilibrium with the bath. 
\subsection{Resolved Entropy Production}
By \eqref{fullentropy}, $\Sigma_{\rm full}$ is the mean of the endpoint term plus the mean bath term. The endpoint term is $\varsigma_{\rm end}=(f-i)\ln(1/\zeta_\epsilon)$. The initial outcome $i$ is read in the eigenbasis of $\rho_{\rm ss}$, so $\langle i\rangle=\nu_\epsilon-\tfrac12$. The final outcome $f$ is read in the basis $\{\Theta^{-1}|\tilde\psi_f\rangle\}$, so $\langle f\rangle$ is the Williamson number of the reverse stationary state evaluated in $\Theta\rho_{\rm ss}\Theta^{-1}$. That Gaussian expectation is computed in Appendix A, Sec. 6. With $\tilde X_{\rm ss}$ the covariance \eqref{covariance} with $g\to-g$ and $E=\mathrm{diag}(1,-1)$,
\begin{align}
\label{BMatrix}
    B_\epsilon&=\langle\varsigma_{\rm end}\rangle
\notag \\ &=\frac{\nu_\epsilon}{2}\ln\frac1{\zeta_\epsilon}\,\operatorname{tr}\bigl(\tilde X_{\rm ss}^{-1}EX_{\rm ss}E-\mathbb I\bigr)
\notag \\&=\frac{8\gamma^2g^2}{\mathfrak a\,\mathfrak d}\,\nu_\epsilon\ln\frac1{\zeta_\epsilon}.
\end{align}
$B_\epsilon$ is positive and independent of $\mathcal T$. It depends continuously on $\epsilon$ on $[0,1)$, with a finite vacuum value $B_0$. It is nonzero because the final measurement basis is the time reverse of the reverse experiment’s reference basis, not the eigenbasis of the forward final state. It is therefore a property of the chosen endpoint convention.

Combining \eqref{fullentropy}, \eqref{JCurrent}, and \eqref{BMatrix} with $(1+\epsilon)/(1-\epsilon)=\coth(\hbar\omega_c/2k_BT_b)$ and $\ln(1/\epsilon)=\hbar\omega_c/k_BT_b$, gives the result
\begin{equation}
\label{SigmaFullExact}
    \Sigma_{\rm full}(\mathcal T,T_b)
=\frac{\mathcal T\dot Q_0}{k_BT_b}\coth\frac{\hbar\omega_c}{2k_BT_b}+B_\epsilon .
\end{equation}
Since $\coth x-1=O(e^{-2x})$ and $B_\epsilon\to B_0$, at fixed observation time this becomes
\begin{equation}
\label{SigmaFull}
    \Sigma_{\rm full}(\mathcal T,T_b)=\frac{\mathcal T\dot Q_0}{k_BT_b}+B_0+o(1),
\qquad T_b\downarrow0 .
\end{equation}
Equation \eqref{SigmaFull} is the microscopic side of the main result. The resolved entropy production consists of the Clausius entropy of the vacuum heat current plus a bounded endpoint term. The leading term grows as $1/T_b$ with coefficient $\mathcal T\dot Q_0/k_B$, which is fixed by the drive, detuning, and  damping. The exponentially small correction from the coth factor is thermal stimulation of the current, which vanishes faster than any power of $T_b$. The temperature dependence of the leading term therefore comes from the factor $\hbar\omega_c/k_BT_b$ that each emitted quantum carries in the likelihood ratio. In the opposite limit $k_BT_b\gg\hbar\omega_c$, $x\coth x\to1$, and the resolved rate $\dot\Sigma_{\rm full}=2\dot J_0\,x\coth x$, with $x=\hbar\omega_c/2k_BT_b$, approaches the temperature-independent value $2\dot J_0$. 

At exactly zero temperature, the divergence follows directly from support. Let \(\mathcal E=\{N_{\mathrm e}\ge1\}\). Since \(\mathbb E_F[N_{\mathrm e}]=\mathcal T\dot J_0>0\), this event has positive forward probability. Its aligned reversal requires at least one absorption from the vacuum, which is forbidden. Consequently,
\begin{equation}
\label{zeroT}
    P_0(\mathcal E)>0,
\quad
Q_0(\mathcal E)=0,
\quad
\Sigma_{\mathrm{full}}(\mathcal T,0)=\infty.
\end{equation}
Thus a finite heat current coexists with infinite resolved irreversibility. The stationary dynamics remains regular; the singularity lies in the disappearance of reverse energy exchanges. This fixes the coefficient relevant to the hidden-entropy law since once the specified counter’s observed relative entropy is shown to remain bounded, the entire leading \(1/T_b\) contribution remains hidden. The following detector analysis establishes that boundedness and the retention of a strictly positive observed arrow of time.

\section{Central Result: Hidden Irreversibility}
\begin{figure*}
\includegraphics{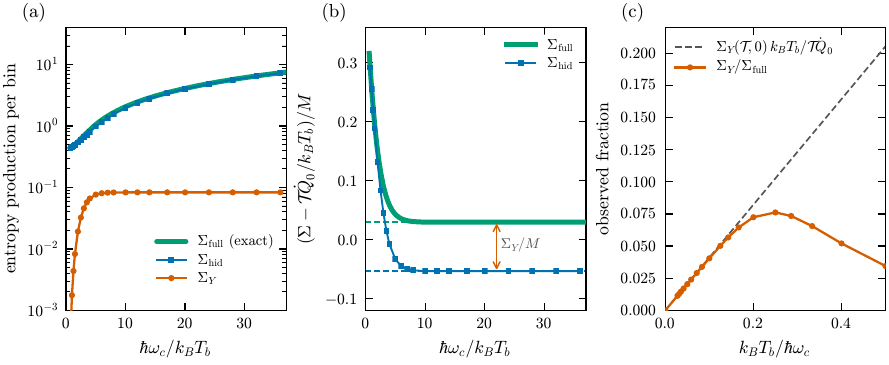}% Here is how to import EPS art
\caption{\label{fig:wide}Resolved, observed and hidden entropy production over a window of $M=8$ bins, for $\Delta=\gamma$, $g=0.6\gamma$, $\gamma\delta=1$ and a counter with $\eta=1$, $\vartheta=0.8$ and $r=0.05$. Values are per bin. $\Sigma_{\rm full}$ is the exact result \eqref{SigmaFullExact}. $\Sigma_Y$ is obtained by exact enumeration of all $2^M$ detector words, and $\Sigma_{\rm hid}=\Sigma_{\rm full}-\Sigma_Y$. (a) As the bath cools, $\Sigma_{\rm full}$ and $\Sigma_{\rm hid}$ grow without bound, while $\Sigma_Y$ saturates at its vacuum value $\Sigma_Y(\mathcal T,0)/M\approx0.083$. (b) The same quantities with the Clausius term $\mathcal T\dot Q_0/k_BT_b$ subtracted. The remainders converge to the constants $B_0/M$ and $[B_0-\Sigma_Y(\mathcal T,0)]/M$ of Eq.~\eqref{HiddenRemainder} (dashed lines), and their separation is $\Sigma_Y/M$. (c) The observed fraction $\Sigma_Y/\Sigma_{\rm full}$ vanishes linearly as $T_b\to0$, along the asymptote $\Sigma_Y(\mathcal T,0)\,k_BT_b/\mathcal T\dot Q_0$ (dashed).}
\end{figure*}
In this section, we discuss the central result. We define a realistic photon counter and the observed forward and reverse laws it produces, and split the resolved entropy production into an observed and a hidden part. We then show that the observed part stays finite and strictly positive at vacuum while the hidden part carries the full divergence of Sec. III, so that the hidden irreversibility grows as the Clausius entropy of the vacuum heat current.
\subsection{The Counter}
The window $[0,\mathcal T)$ is divided into $M$ bins $I_j=[(j-1)\delta,j\delta)$ with $\mathcal T=M\delta$, and the counter reports one bit $y_j\in\{0,1\}$ per bin. Let $n_j(\Gamma)$ be the number of emissions in bin $j$, that is, collisions in which the ancilla gains a quantum. Three imperfections act independently of the system and of each other. Each emitted photon is registered independently with probability $\eta$, which includes the outcoupling fraction $\gamma_{\rm ext}/\gamma$ and the detector’s quantum efficiency. Each bin is live independently with probability $\vartheta$, and the counter does not record which bins were dead. We call this gating. This models a randomly chopped or duty-cycled detector. Each bin independently produces a false click (which we will call a dark count) with probability $r=1-e^{-\nu_d\delta}$, where $\nu_d$ is the dark-count rate. A bin is flagged if it is live and at least one photon is collected, or if a dark count occurs. Given the fine record, the probability of an empty bin is
\begin{align}
\label{kernelbin}
    k(0\mid n)&=(1-r)\bigl[1-\vartheta+\vartheta(1-\eta)^n\bigr], \notag \\
\quad
k(1\mid n)&=1-k(0\mid n),
\end{align}
and the detector is the Markov kernel
\begin{equation}
\label{kernel}
    K(y\mid\Gamma)=\prod_{j=1}^M k\bigl(y_j\mid n_j(\Gamma)\bigr).
\end{equation}
The output is therefore a classical post-processing of the resolved record. The fine record and the detector word are produced together by one joint instrument, and the detector adds no information about the system beyond what the fine record already contains. From here onwards, we will take $0<r<1$, $0<\eta\le1$ and $0<\vartheta\le1$, and we will state explicitly where $\vartheta<1$ is needed.
\subsection{Observed Forward and Reversed Laws}
The observed laws are obtained by applying the same kernel to the forward law and to the aligned reverse law of Sec. II.C,
\begin{align}
    p_Y(y)&=\sum_\Gamma K(y\mid\Gamma)\,P_\epsilon(\Gamma), \notag \\
    q_Y(y)&=\sum_\Gamma K(y\mid\Gamma)\,Q_\epsilon(\Gamma).
\end{align}
Both laws thus come from one coarse-graining of the same aligned microscopic records. Physically, $Q_\epsilon(\Gamma)=\tilde P_\epsilon(\tilde\Gamma)$ and alignment maps forward emissions to reverse absorptions with time inverted. Applying $K$ to $Q_\epsilon$ is therefore equivalent to running the reverse experiment with a counter of the same parameters placed on the absorption channel. In the collision realization, that counter flags ancillas that leave with fewer quanta than they arrived with. By \eqref{recordlaw}, the reverse experiment can be replaced by the forward one
\begin{equation}
\label{qYpa}
    q_Y(y)=p_{\rm a}(y^{\rm rev}),
\end{equation}
where $p_{\rm a}$ is the law of the same counter applied to forward absorptions and $y^{\rm rev}_j=y_{M+1-j}$. The observed and hidden irreversibilities are
\begin{align}
\label{observedhidden}
    \Sigma_Y(\mathcal T,T_b)&=D_{\rm KL}(p_Y\Vert q_Y), \notag \\
    \Sigma_{\rm hid}(\mathcal T,T_b)&=\sum_y p_Y(y)\,D_{\rm KL}\bigl(P_\epsilon(\cdot\mid y)\,\Vert\,Q_\epsilon(\cdot\mid y)\bigr),
\end{align}
where $P_\epsilon(\cdot\mid y)$ and $Q_\epsilon(\cdot\mid y)$ are the conditional laws of the fine record given the output under the joint laws $P_\epsilon K$ and $Q_\epsilon K$. The chain rule for relative entropy applied to these joint laws gives
\begin{equation}
\label{chainrule}
    \Sigma_{\rm full}(\mathcal T,T_b)=\Sigma_Y(\mathcal T,T_b)+\Sigma_{\rm hid}(\mathcal T,T_b).
\end{equation}
The identity requires that the same kernel act on both laws. Without that, the joint relative entropy would not reduce to $\Sigma_{\rm full}$. Both terms are nonnegative, so $\Sigma_Y\le\Sigma_{\rm full}$, which is data processing.
\subsection{The Vacuum Singularity}
\begin{figure*}
\includegraphics{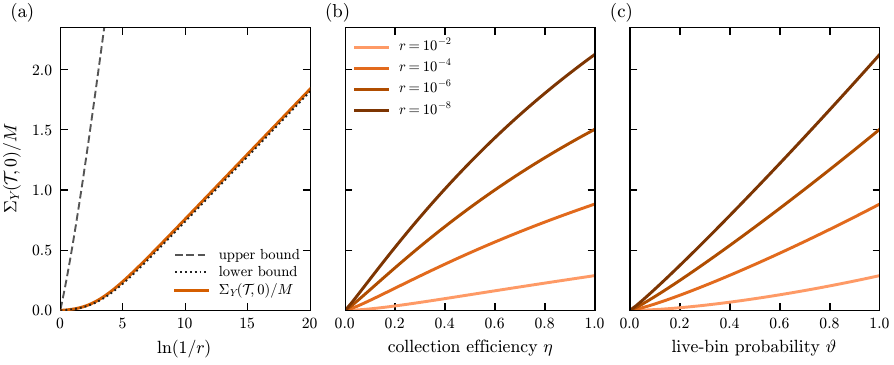}% Here is how to import EPS art
\caption{\label{fig:wide}Observed entropy production at vacuum over a window of $M=8$ bins, for $\Delta=\gamma$, $g=0.6\gamma$ and $\gamma\delta=1$. Values are per bin, and $\Sigma_Y(\mathcal T,0)$ is obtained by exact enumeration of all $2^M$ detector words against the vacuum reverse law $\mathrm{Bern}(r)^{\otimes M}$. (a) $\Sigma_Y(\mathcal T,0)/M$ against $\ln(1/r)$ for the counter of Fig.~2 ($\eta=1$, $\vartheta=0.8$), with the lower (dotted) and upper (dashed) bounds of Eq.~\eqref{thrm1}. The observed entropy production grows without bound as $r\to0$, with slope equal to the click probability of a counter with no dark counts, $\vartheta[1-G_\delta(1-\eta)]\approx0.108$. The upper bound grows with slope $\vartheta$. (b) Dependence on the collection efficiency $\eta$ at $\vartheta=1$, and (c) on the live-bin probability $\vartheta$ at $\eta=1$, for $r=10^{-2},10^{-4},10^{-6},10^{-8}$ (light to dark). Photon loss and gating reduce $\Sigma_Y$, which vanishes only when no photon can be registered, but leave it finite for every $r>0$. Each hundredfold reduction of $r$ raises $\Sigma_Y$ by nearly the same amount, so it diverges as $r\to0$ for every $\eta>0$ and $\vartheta>0$.}
\end{figure*}
At vacuum every ancilla arrives in $|0\rangle$, so any record with an emission has some $k_n'=1$. Its aligned reverse weight contains the factor $\rho_B(1)=0$ by \eqref{loglikelihood}, so
\begin{equation}
\label{mismatch}
    Q_0(\Gamma)=0 \quad \forall N_{\rm e}(\Gamma)\ge1 .
\end{equation}
Under the forward law, $\langle N_{\rm e}\rangle=\mathcal T\dot J_0>0$ by \eqref{JCurrent} meaning such records have positive probability. $P_0$ is therefore not absolutely continuous with respect to $Q_0$, and
\begin{equation}
    \Sigma_{\rm full}(\mathcal T,0)=\infty \quad \forall\; \mathcal{T} >0
\end{equation}
which is consistent with the divergence that we saw in \eqref{SigmaFull}. The divergence still persists even upon conditioning on every detector output. When $\vartheta<1$, every bin has $k(0\mid n)\ge(1-r)(1-\vartheta)$ and $k(1\mid n)\ge r$, so $K(y\mid\Gamma)\ge m^M$ for all $y$ and $\Gamma$, with
\begin{equation}
    m=\min\{r,\,(1-r)(1-\vartheta)\}>0 .
\end{equation}
This floor gives records with emissions a positive conditional weight under the forward law for every output, while the reverse law gives them zero, so $\Sigma_{\rm hid}(\mathcal T,0)=\infty$ conditionally on every detector output (Appendix B).
\subsection{Detector Bounds}
\textbf{Theorem 1: Detector Bounds.} Fix $\mathcal T=M\delta$ and detector parameters with $0<r<1$, $0<\eta\leq1$ and $0<\vartheta\leq1$.
\\
\\
(i) At vacuum,
\begin{align}
\label{thrm1}
    0&<M\,d_{\rm B}(p_*\Vert r)\notag \\ &\le \Sigma_Y(\mathcal T,0)\notag \\ &\le M\,d_{\rm B}\bigl(r+(1-r)\vartheta\,\Vert\,r\bigr)<\infty,
\end{align}
where $d_{\rm B}(p\Vert r)=p\ln(p/r)+(1-p)\ln[(1-p)/(1-r)]$ and $p_*$ is the stationary probability that a bin is flagged.
\\
\\
(ii) At every temperature when $\vartheta <1$,
\begin{equation}
\label{thrm2}
    0\le\Sigma_Y(\mathcal T,T_b)\le M\ln\frac1m .
\end{equation}
\\
The proof of Theorem 1 is given in Appendix B. The three detector imperfections that we have mentioned before now have clear effects on \eqref{thrm1} and \eqref{thrm2}. If the dark-count probability were zero, the aligned reverse word at vacuum would be zero while the forward counter flags bins with probability $p_*>0$ (Fig.~3). The observed laws would then inherit the support mismatch \eqref{mismatch}, and $\Sigma_Y(\mathcal T,0)$ would be infinite. This holds for every $\eta>0$, $\vartheta>0$, and bin width. Missed photons, gating, and binning thin the forward clicks but never produce a reverse click, so none of them can conceal the singularity. Dark counts, however, do because every word has probability at least $\min\{r,1-r\}^M$ under the vacuum reverse law. What remains distinguishable is the excess of the forward click probability over the dark level. That excess is set by the emission statistics and vanishes only if the counter cannot register photons at all. The gate is not needed for \eqref{thrm1}. It supplies the uniform floor $m$, which yields a temperature-independent bound \eqref{thrm2} and makes the hidden part infinite for every output. Without it, $k(0\mid n)\to0$ as $n\to\infty$ and no uniform floor exists. 

\subsection{Hidden Irreversibility at Low Temperatures}
Combining the chain rule \eqref{chainrule}, the exact resolved entropy production \eqref{SigmaFull} and the bound \eqref{thrm2} gives 
\begin{align}
    \frac{\mathcal T\dot Q_0}{k_BT_b}\coth\frac{\hbar\omega_c}{2k_BT_b}&+B_\epsilon-M\ln\frac1m
\notag \\ &\le\Sigma_{\rm hid}(\mathcal T,T_b)\notag \\ &\le
\frac{\mathcal T\dot Q_0}{k_BT_b}\coth\frac{\hbar\omega_c}{2k_BT_b}+B_\epsilon .
\end{align}
for all $T_b>0. $ As $T_b\to0$ at fixed $\mathcal T$ and fixed detector parameters, the coth factor approaches one exponentially fast and $B_\epsilon\to B_0$, giving
\begin{equation}
    \Sigma_{\rm hid}(\mathcal T,T_b)=\frac{\mathcal T\dot Q_0}{k_BT_b}+O(1),
\qquad T_b\downarrow0,
\end{equation}
with the bounded remainder between $B_0-M\ln(1/m)$ and $B_0$. These results are visualized in Fig.~2. Continuity of $\Sigma_Y$ at vacuum (Appendix B) identifies the remainder as
\begin{equation}
\label{HiddenRemainder}
    \Sigma_{\rm hid}(\mathcal T,T_b)=\frac{\mathcal T\dot Q_0}{k_BT_b}+\bigl[B_0-\Sigma_Y(\mathcal T,0)\bigr]+o(1).
\end{equation}
Equations \eqref{thrm1} to \eqref{HiddenRemainder} are the central results of this paper. They give a precise size to the irreversibility a realistic observation conceals, and they separate the parts fixed by physics from the parts constrained by the instrument.

The hidden part is, to leading order, the entire Clausius entropy of the heat the oscillator delivers to the bath. It does not depend on the detector; indeed, no choice of $\eta$, $\vartheta$, $\delta$ or $r$ in the admissible range changes the rate at which irreversibility is hidden. The fraction of the microscopic irreversibility the counter registers, $\Sigma_Y/\Sigma_{\rm full}$, therefore vanishes linearly in $T_b$. The counter still sees a definite arrow of time, but the amount it sees is fixed by how often it clicks spuriously, not by how much heat the system dissipates. 

\section{Conclusion}
We have constructed a driven dissipative quantum system whose microscopic irreversibility is infinite at vacuum, with an imperfect counter that conceals a divergent contribution to the microscopic irreversibility. The collision realization specifies the bath energy exchanges and their physical reverse, while the detector defines an explicit coarse-graining of the resolved record. At vacuum, the observed relative entropy is finite and strictly positive, whereas the microscopic and hidden relative entropies are infinite. At small positive bath temperature, for fixed observation duration and detector parameters, the hidden entropy diverges inversely with the bath temperature $T_b$. The mechanism is the disappearance of reverse processes. The pump sustains emission into a bath that, at vacuum, cannot supply an excitation for the reversed absorption. Consequently, a set of forward histories with positive probability has zero reverse probability. Dark counts conceal the distinction because a reported click can arise even when the corresponding physical exchange is absent. Nevertheless, emissions still increase the forward click probability above the reverse dark-count baseline, preserving a finite arrow of time.

The distinction between missed events and false positives is essential to this result. Photon loss, gating, and finite time bins reduce the number of detected emissions, but any registered emission still has zero reverse probability at vacuum when false positives are absent. These imperfections therefore preserve the divergence whenever the counter retains a nonzero probability of detecting an emission. Dark counts conceal the support mismatch by allowing clicks even when the corresponding physical exchange is forbidden. For the specified detector, the observed irreversibility remains finite and strictly positive, while the hidden contribution satisfies \(\Sigma_{\mathrm{hid}}(\mathcal T,T_b)=\mathcal T\dot Q_0/(k_BT_b)+O(1)\) as \(T_b\to0\), at fixed observation duration and detector parameters. The vacuum heat current thus fixes the leading divergence independently of the detector settings within the admissible range. 

This separates physical irreversibility from the evidence available to an observer. Each system–ancilla collision admits an inverse unitary, but the prescribed reverse experiment uses fresh thermal ancillas. At vacuum, those ancillas cannot supply the excitations required to reverse an emission. The asymmetry thus depends on the physical preparation of the environment, while its visibility depends on the measurement. Infinite relative entropy also does not mean that every individual history identifies time’s direction with certainty because histories without exchanges can remain ambiguous. 

The broader consequence is a limit on thermodynamic inference from partial observations. Coarse-grained entropy production provides a lower bound on resolved irreversibility, as established in stochastic thermodynamics and imperfect quantum monitoring. Our construction shows that this bound can remain finite and informative while the unresolved contribution diverges. A finite measured entropy production therefore cannot establish that the underlying thermodynamic entropy production is finite, and a small measured value cannot establish near reversibility. Such conclusions require additional knowledge of unresolved exchanges and detector response. The concealment concerns model-free inference from the records. An observer who knows the bath temperature, assumes local detailed balance, and calibrates the collection efficiency can estimate the net emission rate from the click rate and recover the leading Clausius term by thermodynamic reasoning. The record-level relative entropy cannot, because it charges each click at most the cost of a dark count rather than the cost of a forbidden absorption. Likewise, regular stationary states and smooth low-order observables need not imply regular trajectory thermodynamics.

Potential future work could establish long-time entropy-production rates or extend the analysis to interacting and multimode systems. In cosmology, the framework could help determine whether eliminating momentum variables or inaccessible field modes in stochastic inflation conceals irreversibility. In quantum information, extending thermodynamic inference under imperfect monitoring to quantum memories and feedback-based error correction could clarify how much entropy production remains invisible in accessible measurement records. These applications could guide the design of measurements that distinguish nearly reversible dynamics from substantial irreversibility concealed by incomplete observation.

\appendix 
\section{Stationary Gaussian State}
This appendix derives the stationary state of the forward generator $\mathcal L_\epsilon$ used in Sec. III. We obtain its covariance matrix, mean occupation and symplectic eigenvalue, and show that these fix its spectrum, Eq.~\eqref{zeta}, and keep it strictly mixed at vacuum whenever $g>0$.
\\
\\
\textbf{1. Moment equations}
\\
For an operator $O$, the generator \eqref{MasterEq} gives
\begin{align}
    \frac{d\langle O\rangle}{dt}=\frac{i}{\hbar}\langle[H_0,O]\rangle+\gamma(\bar n+1)&\Bigl\langle a^\dagger Oa-\tfrac12\{a^\dagger a,O\}\Bigr\rangle +\notag \\ &\gamma\bar n\Bigl\langle aOa^\dagger-\tfrac12\{aa^\dagger,O\}\Bigr\rangle .
\end{align}
For $O=a$, using $[a^\dagger a,a]=-a$ and $[a^{\dagger2},a]=-2a^\dagger$, the Hamiltonian contributes $-i\Delta\langle a\rangle+g\langle a^\dagger\rangle$. The two dissipators together contribute $-\tfrac{\gamma}{2}\langle a\rangle$, independent of $\bar n$. This gives the mean-field equation \eqref{meanfield}. In terms of the quadratures $R=(Q,P)$, with $Q=(a+a^\dagger)/\sqrt2$ and $P=(a-a^\dagger)/(i\sqrt2)$, the means obey $d\langle R\rangle/dt=A\langle R\rangle$, with
\begin{equation}
    A=\begin{bmatrix} g-\gamma/2 & \Delta\\ -\Delta & -g-\gamma/2\end{bmatrix}.
\end{equation}
Its eigenvalues are $-\gamma/2\pm\sqrt{g^2-\Delta^2}$, so $A$ is stable (all eigenvalues have negative real part) exactly when $\mathfrak d>0$. Below threshold, the means therefore decay to zero and the stationary state is centred. Applying (A1) to the symmetrized second moments gives
\begin{align}
    \frac{dX}{dt}=AX+XA^{\mathsf T}&+D,
\quad
X_{jk}=\tfrac12\langle\{R_j,R_k\}\rangle, \notag \\ 
&D=\gamma\Bigl(\bar n+\tfrac12\Bigr)\mathbb I .
\end{align}
In this convention the vacuum has $X=\mathbb I/2$. As a check, at $g=\Delta=0$ Eq. (A3) relaxes to $X=(\bar n+\tfrac12)\mathbb I$, the thermal state of the bath.
\\
\\
\textbf{2. Stationary covariance}
\\
Write $X=\begin{bmatrix}x&z\\ z&y\end{bmatrix}$ and $c=2\bar n+1$. The stationary condition $AX+XA^{\mathsf T}+D=0$ has three independent components:
\begin{align}
    (2g-\gamma)\,x+2\Delta z+\tfrac{\gamma c}{2}&=0, &&(QQ)\notag \\
-(2g+\gamma)\,y-2\Delta z+\tfrac{\gamma c}{2}&=0, &&(PP)\\
-\gamma z+\Delta(y-x)&=0. &&(QP) \notag
\end{align}
Introduce $S=x+y$ and $s=x-y$. The $(QP)$ equation gives $z=-\Delta s/\gamma$. Adding and subtracting the first two equations gives
\begin{equation}
    2gs-\gamma S+\gamma c=0,
\quad
2gS-\gamma s+4\Delta z=0 .
\end{equation}
Substituting $z$ into the second equation of (A5) gives $2gS=s\,\mathfrak a/\gamma$, with $\mathfrak a=\gamma^2+4\Delta^2$, so $s=2\gamma gS/\mathfrak a$. The first equation then reads $S(\mathfrak a-4g^2)/\mathfrak a=c$. Since $\mathfrak a-4g^2=\mathfrak d$, this gives
\begin{equation}
    S=\frac{c\,\mathfrak a}{\mathfrak d},
\quad
s=\frac{2\gamma g\,c}{\mathfrak d},
\quad
z=-\frac{2\Delta g\,c}{\mathfrak d},
\end{equation}
and hence
\begin{equation}
    X_{\rm ss}=\frac{2\bar n+1}{2\mathfrak d}
\begin{bmatrix}\mathfrak a+2\gamma g & -4\Delta g\\ -4\Delta g & \mathfrak a-2\gamma g\end{bmatrix},
\end{equation}
which is Eq.~\eqref{covariance}. For $\mathfrak d>0$, $A$ is stable, so the Lyapunov equation has exactly this one solution, and $X(t)\to X_{\rm ss}$ from every initial state. The pump makes the $Q$ variance larger than the $P$ variance by $2\gamma g\,c/\mathfrak d$. The detuning couples the two quadratures through $z\propto\Delta g$, so at zero detuning the ellipse is aligned with the $Q$ and $P$ axes.
\\
\\
\textbf{3. Occupation and anomalous moment}
\\
Since $\langle a^\dagger a\rangle=\tfrac12(x+y)-\tfrac12=\tfrac12(S-1)$ and $\mathfrak a=\mathfrak d+4g^2$,
\begin{equation}
    n_{\rm ss}=\frac{c\,\mathfrak a-\mathfrak d}{2\mathfrak d}=\bar n+(2\bar n+1)\frac{2g^2}{\mathfrak d},
\end{equation}
which is Eq.~\eqref{nss}. Similarly, $\langle a^2\rangle=\tfrac12(x-y+2iz)$, which gives
\begin{equation}
    \langle a^2\rangle=(2\bar n+1)\,\frac{g(\gamma-2i\Delta)}{\mathfrak d}.
\end{equation}
The real part enters the pump balance of Sec. III.B: the pump changes the occupation at rate $2g\,{\rm Re}\langle a^2\rangle=2\gamma g^2(2\bar n+1)/\mathfrak d=\dot J(\epsilon)$. At vacuum, $n_{\rm ss}=2g^2/\mathfrak d>0$ and $\langle a^2\rangle\neq0$. The cavity holds a squeezed, populated field even though the bath supplies no quanta.
\\
\\
\textbf{4. Symplectic eigenvalue and spectrum}
\\
The determinant of (A7) is
\begin{align}
    \det X_{\rm ss}&=\Bigl(\frac{c}{2\mathfrak d}\Bigr)^2\Bigl[\mathfrak a^2-4\gamma^2g^2-16\Delta^2g^2\Bigr] \notag \\
&=\Bigl(\frac{c}{2\mathfrak d}\Bigr)^2\mathfrak a\,(\mathfrak a-4g^2) \notag \\
&=\Bigl(\frac{c}{2}\Bigr)^2\frac{\mathfrak a}{\mathfrak d},
\end{align}
so the symplectic eigenvalue is
\begin{equation}
    \nu_\epsilon=\sqrt{\det X_{\rm ss}}=\Bigl(\bar n+\tfrac12\Bigr)\sqrt{\frac{\mathfrak a}{\mathfrak d}},
\end{equation}
which is Eq.~\eqref{nu}. For one mode, $X_{\rm ss}/\nu_\epsilon$ is a positive definite $2\times2$ matrix with unit determinant, and every such matrix is symplectic. Its positive square root $S_\epsilon=(X_{\rm ss}/\nu_\epsilon)^{1/2}$ is therefore symplectic, and
\begin{equation}
    X_{\rm ss}=\nu_\epsilon\,S_\epsilon S_\epsilon^{\mathsf T}.
\end{equation}
Let $U_{S_\epsilon}$ be the Gaussian unitary that implements $R\mapsto S_\epsilon R$, and $\tau_\nu$ the thermal state with covariance $\nu\mathbb I$. Because a centred Gaussian state is fixed by its covariance, Eq. (A12) gives
\begin{equation}
    \rho_{\rm ss}=U_{S_\epsilon}\,\tau_{\nu_\epsilon}\,U_{S_\epsilon}^\dagger .
\end{equation}
The thermal state $\tau_\nu$ has mean occupation $\nu-\tfrac12$ and geometric populations, so
\begin{align}
    \rho_{\rm ss}=\sum_{j\ge0}p_j\,|\psi_j\rangle\langle\psi_j|,&
\qquad
p_j=(1-\zeta_\epsilon)\zeta_\epsilon^{\,j}, \notag \\ 
\zeta_\epsilon=\frac{\nu_\epsilon-\tfrac12}{\nu_\epsilon+\tfrac12}, &
\qquad
|\psi_j\rangle=U_{S_\epsilon}|j\rangle .
\end{align}
This is Eq.~\eqref{density}, with Eq.~\eqref{zeta}. The spectrum is nondegenerate whenever $0<\zeta_\epsilon<1$, so the eigenbasis $\{|\psi_j\rangle\}$ used for the endpoint measurements is unique up to phases. The endpoint term of Eq.~\eqref{endpoint} involves
\begin{equation}
    \ln\frac1{\zeta_\epsilon}=\ln\frac{\nu_\epsilon+\tfrac12}{\nu_\epsilon-\tfrac12},
\end{equation}
which is the effective inverse temperature of the stationary state in its Williamson basis.
\\
\\
\textbf{5. Mixedness at vacuum}
\\
The purity of a one-mode Gaussian state is $\operatorname{tr}\rho^2=1/(2\sqrt{\det X})$, so
\begin{equation}
    \operatorname{tr}\rho_{\rm ss}^2=\frac{1}{2\nu_\epsilon}.
\end{equation}
At vacuum, $\bar n=0$ and (A11) becomes $\nu_0=\tfrac12\sqrt{\mathfrak a/\mathfrak d}$. Since $\mathfrak a-\mathfrak d=4g^2$, we have $\nu_0>\tfrac12$ if and only if $g\neq0$. For every $g>0$ below threshold, therefore,
$
\operatorname{tr}\rho_{\rm ss}^2<1,
\quad
0<\zeta_0<1,
\quad
p_j>0\ \text{for all } j .
$
The stationary state is strictly mixed and faithful (full rank), even though the bath is in a pure state. At positive temperature, $\nu_\epsilon=(2\bar n+1)\nu_0$ only increases, so $0<\zeta_\epsilon<1$ holds at every temperature. This is the faithfulness used in Sec. II.C. It is what makes $\varsigma_{\rm end}$ finite on every record and keeps the boundary term $B_\epsilon$ finite at $T_b=0$.

The physical reason is that the damping and the pump have no common pure fixed point. With the single jump operator $L_{\rm e}=\sqrt\gamma\,a$ at vacuum, a pure stationary state would have to be an eigenstate of $a$, which is a coherent state. Since the stationary mean vanishes, that coherent state would be the vacuum $|0\rangle$. But the pump term moves the vacuum: $H_0|0\rangle$ has a $|2\rangle$ component of amplitude $i\hbar g/\sqrt2$. Down-conversion keeps creating pairs that damping removes one quantum at a time, and the balance between them is a mixed squeezed thermal state.

For the parameters used in the figures, $\Delta=\gamma$ and $g=0.6\gamma$, so $\mathfrak a=5\gamma^2$ and $\mathfrak d=3.56\gamma^2$. At vacuum this gives $n_{\rm ss}\approx0.202$, $\nu_0\approx0.593$, $\zeta_0\approx0.085$ and purity $\approx0.84$.
\\
\\
\textbf{6. The endpoint term $B_\epsilon$}
\\
By \eqref{endpoint}, $\varsigma_{\rm end}=(f-i)\ln(1/\zeta_\epsilon)$, so $B_\epsilon=\ln(1/\zeta_\epsilon)\,(\langle f\rangle-\langle i\rangle)$. Both means are expectations of Williamson number operators. For a centered Gaussian state with covariance $Y=\nu\,S S^{\mathsf T}$, the number operator in its eigenbasis, $N_Y=U_S\,a^\dagger a\,U_S^\dagger$, is the quadratic form
\begin{equation}
    N_Y=\frac{\nu}{2}\,R^{\mathsf T}Y^{-1}R-\frac12 ,
\end{equation}
and its mean in any centered Gaussian state with covariance $Z$ is
\begin{equation}
    \langle N_Y\rangle_Z=\frac{\nu}{2}\operatorname{tr}\bigl(Y^{-1}Z\bigr)-\frac12 .
\end{equation}
The initial outcome is read in the eigenbasis of $\rho_{\rm ss}$, so $\langle i\rangle=\langle N_{X_{\rm ss}}\rangle_{X_{\rm ss}}=\nu_\epsilon-\tfrac12$. The final outcome is read in the basis $\{\Theta^{-1}|\tilde\psi_f\rangle\}$ on the state $\rho_{\rm ss}$. Its probabilities are $\langle\tilde\psi_f|\Theta\rho_{\rm ss}\Theta^{-1}|\tilde\psi_f\rangle$, so $\langle f\rangle$ is the mean of the reverse state’s number operator $N_{\tilde X_{\rm ss}}$ in the state $\Theta\rho_{\rm ss}\Theta^{-1}$. Then we have two covariances. First, time reversal sends $(Q,P)\mapsto(Q,-P)$, so $\Theta\rho_{\rm ss}\Theta^{-1}$ has covariance $EX_{\rm ss}E$, with $E={\rm diag}(1,-1)$. Second, the quarter turn $U_q$ rotates the quadratures by $90^\circ$. This swaps the diagonal entries of (A7) and flips the sign of the off-diagonal one, so $\tilde X_{\rm ss}$ is (A7) with $g\to-g$. Its symplectic eigenvalue is also $\nu_\epsilon$. Hence
\begin{equation}
    B_\epsilon=\frac{\nu_\epsilon}{2}\ln\frac1{\zeta_\epsilon}\operatorname{tr}\bigl(\tilde X_{\rm ss}^{-1}EX_{\rm ss}E-\mathbb I\bigr).
\end{equation}
For $2\times2$ matrices, $\tilde X_{\rm ss}^{-1}=\operatorname{adj}(\tilde X_{\rm ss})/\nu_\epsilon^2$, and here $\operatorname{adj}(\tilde X_{\rm ss})=X_{\rm ss}$. With $c=2\bar n+1$, this gives
\begin{align}
    \operatorname{tr}\bigl(\tilde X_{\rm ss}^{-1}EX_{\rm ss}E\bigr)
&
=\frac{2\mathfrak a^2+8\gamma^2g^2-32\Delta^2g^2}{\mathfrak a\mathfrak d}.
\end{align}
Subtracting 2 and using $\mathfrak a-\mathfrak d=4g^2$ and $\mathfrak a+\gamma^2-4\Delta^2=2\gamma^2$,
\begin{align}
    \operatorname{tr}\bigl(\tilde X_{\rm ss}^{-1}EX_{\rm ss}E-\mathbb I\bigr)
&=\frac{8g^2(\mathfrak a+\gamma^2-4\Delta^2)}{\mathfrak a\mathfrak d} \notag \\ 
&=\frac{16\gamma^2g^2}{\mathfrak a\mathfrak d},
\end{align}
and therefore
\begin{equation}
    B_\epsilon=\frac{8\gamma^2g^2}{\mathfrak a\,\mathfrak d}\,\nu_\epsilon\ln\frac1{\zeta_\epsilon},
\end{equation}
which is Eq.~\eqref{BMatrix}. $B_\epsilon$ is positive for $g>0$ and independent of $\mathcal T$. It is continuous on $\epsilon\in[0,1)$, because $\nu_\epsilon$ is continuous and $\nu_\epsilon\ge\nu_0>\tfrac12$ keeps $\ln(1/\zeta_\epsilon)$ finite. It vanishes at $g=0$, where $EX_{\rm ss}E=\tilde X_{\rm ss}$ and the final measurement basis coincides with the eigenbasis of the forward final state. For the parameters of the figures, $B_0\approx0.2367$.

\section{Proof of Theorem 1}
We will assume $\mathcal T=M\delta$, the detector kernel is \eqref{kernel}, and $n_j(\Gamma)$ is the number of emissions in bin $j$. Sums over $\Gamma$ stand for integrals over records in the continuum limit. By \eqref{kernelbin}, the click probability given $n$ emissions is
\begin{equation}
    k(1\mid n)=r+(1-r)\,\vartheta\bigl[1-(1-\eta)^n\bigr],
\end{equation}
so for every $n\ge0$,
\begin{equation}
    r\;\le\;k(1\mid n)\;\le\;r+(1-r)\vartheta,
\qquad
k(0\mid n)\;\ge\;(1-r)(1-\vartheta).
\end{equation}
\textbf{(i) Bounds at vacuum}
\\
At $T_b=0$ no absorptions occur. A counter applied to forward absorptions therefore flags each bin only through its own dark count, independently with probability $r$. By \eqref{qYpa}, the observed reverse law is
\begin{equation}
    q_Y=\mathrm{Bern}(r)^{\otimes M},
\end{equation}
is the same law for every word and its reversal. Let $\pi_j(y_{<j})=p_Y(y_j=1\mid y_1,\ldots,y_{j-1})$ be the forward conditional click probability. Under (B3) the reverse conditionals are $\mathrm{Bern}(r)$ whatever the past, so the chain rule for relative entropy gives
\begin{equation}
    \Sigma_Y(\mathcal T,0)=\sum_{j=1}^M\Bigl\langle d_{\rm B}\bigl(\pi_j\,\Vert\,r\bigr)\Bigr\rangle_{p_Y}.
\end{equation}
Given the fine record and the earlier outputs, bit $j$ is drawn from $k(\cdot\mid n_j)$ independently of everything else. Hence $\pi_j$ is a conditional average of $k(1\mid n_j)$, and by (B2),
\begin{equation}
    r\le\pi_j\le r+(1-r)\vartheta .
\end{equation}
The function $p\mapsto d_{\rm B}(p\Vert r)$ is convex on $[0,1]$, vanishes at $p=r$, and increases on $[r,1]$.
\\
\\
\textit{Upper bound.} By (B5) and monotonicity, each term in (B4) is at most $d_{\rm B}(r+(1-r)\vartheta\Vert r)$. This is finite because $0<r<1$, and it equals $\ln(1/r)$ at $\vartheta=1$.
\\
\\
\textit{Lower bound.} By Jensen’s inequality, each term is at least $d_{\rm B}(\langle\pi_j\rangle\Vert r)$. Because the forward process is stationary, $\langle\pi_j\rangle=p_Y(y_j=1)=p_*$ for every $j$. Summing over the $M$ bins gives \eqref{thrm1}.
\\
\\
\textit{Strict positivity.} Averaging (B1) over the stationary law gives
\begin{equation}
\label{pstar}
    p_*=r+(1-r)\,\vartheta\bigl[1-G_\delta(1-\eta)\bigr],
\quad
G_\delta(s)=\bigl\langle s^{\,n_j}\bigr\rangle .
\end{equation}
The mean number of emissions per bin is $\delta\dot J_0>0$ by \eqref{JCurrent}, so $P_0(n_j\ge1)>0$. With $\eta>0$ this gives $G_\delta(1-\eta)<1$. Together with $\vartheta>0$ and $r<1$, it follows that $p_*>r$, and therefore $d_{\rm B}(p_*\Vert r)>0$. Part (i) needs only $0<\vartheta\le1$. $\square$
\\
\\
\textbf{(ii) Temperature-uniform bound}
\\
Now let $\vartheta<1$. By (B2), every bin satisfies $k(y_j\mid n)\ge m$ for both values of $y_j$, with $m=\min\{r,(1-r)(1-\vartheta)\}>0$. Hence
\begin{equation}
    K(y\mid\Gamma)\ge m^M\implies 
q_Y(y)=\sum_\Gamma K(y\mid\Gamma)\,Q_\epsilon(\Gamma)\ge m^M 
\end{equation}
for all $y,\Gamma$. Since $\sum_y p_Y\ln p_Y\le0$,
\begin{align}
    \Sigma_Y(\mathcal T,T_b)&=\sum_y p_Y(y)\ln\frac{p_Y(y)}{q_Y(y)}\notag \\ &\le\sum_y p_Y(y)\ln\frac1{q_Y(y)}\notag \\ &\le M\ln\frac1m ,
\end{align}
which is \eqref{thrm2}. The argument uses only the floor (B7), so it holds at every temperature and for any reverse law. $\square$
\\
\\
\textbf{(iii) Conditional divergence at vacuum}
\\
Let $\vartheta<1$ and $\mathcal E=\{N_{\rm e}\ge1\}$. By \eqref{mismatch} and \eqref{zeroT}, $Q_0(\mathcal E)=0$ and $P_0(\mathcal E)>0$. By (B7), every output satisfies $p_Y(y)\ge m^M>0$ and $q_Y(y)\ge m^M>0$, so both conditional laws in \eqref{observedhidden} are defined for every $y$. Under the joint laws $P_0K$ and $Q_0K$,
\begin{align}
    P_0(\mathcal E\mid y)&=\frac{\sum_{\Gamma\in\mathcal E}K(y\mid\Gamma)P_0(\Gamma)}{p_Y(y)}\notag \\ &\ge\frac{m^M P_0(\mathcal E)}{p_Y(y)}>0,
\qquad
Q_0(\mathcal E\mid y)=0 .
\end{align}
For every $y$, therefore, $P_0(\cdot | y)$ is not absolutely continuous with respect to $Q_0(\cdot | y)$, and $D_{\rm KL}(P_0(\cdot | y)\Vert Q_0(\cdot | y))=\infty$. Every output has positive probability, so $\Sigma_{\rm hid}(\mathcal T,0)=\infty$. Whatever word the counter reports, the fine history behind it contains a forbidden emission with positive probability. This is consistent with the chain rule \eqref{chainrule}, since a finite $\Sigma_Y(\mathcal T,0)$ plus an infinite $\Sigma_{\rm hid}(\mathcal T,0)$ gives $\Sigma_{\rm full}(\mathcal T,0)=\infty$. $\square$
\\
\\
\textbf{(iv) Continuity at vacuum}
\\
Fix $M$ and the detector parameters. By \eqref{kernelbin} and \eqref{kernel}, each word probability $p_Y(y)$ is a finite sum, with coefficients depending only on $r$, $\vartheta$ and $\eta$, of terms $\operatorname{tr}[\mathcal S_M\cdots\mathcal S_1\rho_{\rm ss}]$ with each $\mathcal S_j\in\{e^{\delta\mathcal L_\epsilon},e^{\delta(\mathcal L_\epsilon-\eta\mathcal J_{\rm e})}\}$, and $q_Y(y)$ has the same form with $\mathcal J_{\rm a}$ in place of $\mathcal J_{\rm e}$. Every such map sends a centered Gaussian operator to a centered Gaussian operator. Writing the image as $Z\,\hat\rho$ with $\hat\rho$ a normalized Gaussian state of covariance $X$, the emission-tilted map gives
\begin{align}
\label{riccati}
\frac{dX}{dt}&=AX+XA^{\mathsf T}+D-\eta\gamma(\bar n+1)\bigl(X-\tfrac12\mathbb I\bigr)^2, \notag \\
\frac{d\ln Z}{dt}&=-\frac{\eta\gamma(\bar n+1)}{2}\bigl(\operatorname{tr}X-1\bigr),
\end{align}
and the absorption-tilted map gives the same equations with $\gamma(\bar n+1)$ replaced by $\gamma\bar n$ and $X-\tfrac12\mathbb I$, $\operatorname{tr}X-1$ replaced by $X+\tfrac12\mathbb I$, $\operatorname{tr}X+1$. The quadratic terms are negative semidefinite, so $X(t)$ stays below the solution of the linear Lyapunov flow and remains bounded on $[0,\delta]$. The coefficients and the initial covariance $X_{\rm ss}$ depend continuously on $\epsilon\in[0,1)$, so each factor $Z$, and hence each $p_Y(y)$ and $q_Y(y)$, is continuous in $\epsilon$. At vacuum $q_Y=\mathrm{Bern}(r)^{\otimes M}$ is strictly positive on all $2^M$ words, and $p\ln(p/q)$ is continuous wherever $q>0$. The finite sum $\Sigma_Y(\mathcal T,T_b)=\sum_yp_Y\ln(p_Y/q_Y)$ is therefore continuous at $T_b=0$, which gives \eqref{HiddenRemainder}. $\square$

\section{Data Availability}
The observed entropy production in Figs. 2 and 3 was computed exactly for finite windows; there is no sampling. The generator $\mathcal L_\epsilon$ was represented as a matrix superoperator on a truncated Fock space of dimension $N$. The stationary state was obtained as the trace-normalized null vector of that matrix. The bin maps of the counter were built from matrix exponentials:
\begin{equation}
    \mathcal E_0=(1-r)\bigl[(1-\vartheta)e^{\delta\mathcal L_\epsilon}+\vartheta e^{\delta(\mathcal L_\epsilon-\eta\mathcal J_{\rm e})}\bigr],
\quad
\mathcal E_1=e^{\delta\mathcal L_\epsilon}-\mathcal E_0 .
\end{equation}
The absorption-counter maps are the same with $\mathcal J_{\rm a}$ in place of $\mathcal J_{\rm e}$. For $M=8$, all $2^M=256$ forward word probabilities $p_Y(y)=\operatorname{tr}[\mathcal E_{y_M}\cdots\mathcal E_{y_1}\rho_{\rm ss}]$ were enumerated. The aligned reverse law was obtained from the absorption-counter law through Eq.~\eqref{qYpa}; at vacuum it reduces to $\mathrm{Bern}(r)^{\otimes M}$, which was evaluated analytically. $\Sigma_{\rm full}$ was evaluated from the exact expressions \eqref{BMatrix} and \eqref{SigmaFullExact}, and $\Sigma_{\rm hid}$ as the difference $\Sigma_{\rm full}-\Sigma_Y$. The truncation was $N=34$ for $\hbar\omega_c/k_BT_b<1.5$, $N=28$ for $\hbar\omega_c/k_BT_b<3$ and $N=22$ otherwise. Raising $N$ to between 30 and 44 changed $\Sigma_Y/M$ by less than $10^{-8}$. The numerically computed stationary occupation and net emission rate agreed with Eqs.~\eqref{nss} and \eqref{JCurrent} to better than $10^{-5}$. Two further checks were made. The reverse experiment was also simulated directly, with pump phase $\pi$, the reversed generator and its own stationary state; its absorption-word law agreed with Eq.~\eqref{qYpa} to machine precision, confirming the single-experiment identity \eqref{recordlaw}. The single-bin flag probability $p_*$ of Eq.~\eqref{pstar} was recomputed without the Fock basis by integrating the Gaussian equations \eqref{riccati} at vacuum with relative tolerance $10^{-12}$, and agreed with the Fock-space value to within $6\times10^{-10}$ for $\eta\in\{0.25,0.5,1\}$ and $\vartheta\in\{0.8,1\}$. The code can be found in our Github: 
\url{https://github.com/hieutnn/EntropyProduction}

\bibliography{apssamp}

@article{Seifert2005,
  author  = {Seifert, Udo},
  title   = {Entropy production along a stochastic trajectory and an integral fluctuation theorem},
  journal = {Physical Review Letters},
  volume  = {95},
  pages   = {040602},
  year    = {2005}
}

@article{Kawai_2007,
   title={Dissipation: The Phase-Space Perspective},
   volume={98},
   ISSN={1079-7114},
   url={http://dx.doi.org/10.1103/PhysRevLett.98.080602},
   DOI={10.1103/physrevlett.98.080602},
   number={8},
   journal={Physical Review Letters},
   publisher={American Physical Society (APS)},
   author={Kawai, R. and Parrondo, J. M. R. and den Broeck, C. Van},
   year={2007},
   month=Feb }

@article{Parrondo_2009,
   title={Entropy production and the arrow of time},
   volume={11},
   ISSN={1367-2630},
   url={http://dx.doi.org/10.1088/1367-2630/11/7/073008},
   DOI={10.1088/1367-2630/11/7/073008},
   number={7},
   journal={New Journal of Physics},
   publisher={IOP Publishing},
   author={Parrondo, J M R and Broeck, C Van den and Kawai, R},
   year={2009},
   month=July, pages={073008} }

@article{Landi_2021,
   title={Irreversible entropy production: From classical to quantum},
   volume={93},
   ISSN={1539-0756},
   url={http://dx.doi.org/10.1103/RevModPhys.93.035008},
   DOI={10.1103/revmodphys.93.035008},
   number={3},
   journal={Reviews of Modern Physics},
   publisher={American Physical Society (APS)},
   author={Landi, Gabriel T. and Paternostro, Mauro},
   year={2021},
   month=Sept }

@article{Horowitz_2012,
   title={Quantum-trajectory approach to the stochastic thermodynamics of a forced harmonic oscillator},
   volume={85},
   ISSN={1550-2376},
   url={http://dx.doi.org/10.1103/PhysRevE.85.031110},
   DOI={10.1103/physreve.85.031110},
   number={3},
   journal={Physical Review E},
   publisher={American Physical Society (APS)},
   author={Horowitz, Jordan M.},
   year={2012},
   month=Mar }

@article{Horowitz_2013,
   title={Entropy production along nonequilibrium quantum jump trajectories},
   volume={15},
   ISSN={1367-2630},
   url={http://dx.doi.org/10.1088/1367-2630/15/8/085028},
   DOI={10.1088/1367-2630/15/8/085028},
   number={8},
   journal={New Journal of Physics},
   publisher={IOP Publishing},
   author={Horowitz, Jordan M and Parrondo, Juan M R},
   year={2013},
   month=Aug, pages={085028} }

@article{Manzano_2018,
   title={Quantum Fluctuation Theorems for Arbitrary Environments: Adiabatic and Nonadiabatic Entropy Production},
   volume={8},
   ISSN={2160-3308},
   url={http://dx.doi.org/10.1103/PhysRevX.8.031037},
   DOI={10.1103/physrevx.8.031037},
   number={3},
   journal={Physical Review X},
   publisher={American Physical Society (APS)},
   author={Manzano, Gonzalo and Horowitz, Jordan M. and Parrondo, Juan M. R.},
   year={2018},
   month=Aug }

@article{Kawaguchi_2013,
   title={Fluctuation theorem for hidden entropy production},
   volume={88},
   ISSN={1550-2376},
   url={http://dx.doi.org/10.1103/PhysRevE.88.022147},
   DOI={10.1103/physreve.88.022147},
   number={2},
   journal={Physical Review E},
   publisher={American Physical Society (APS)},
   author={Kawaguchi, Kyogo and Nakayama, Yohei},
   year={2013},
   month=Aug }

@article{Esposito_2012,
   title={Stochastic thermodynamics under coarse graining},
   volume={85},
   ISSN={1550-2376},
   url={http://dx.doi.org/10.1103/PhysRevE.85.041125},
   DOI={10.1103/physreve.85.041125},
   number={4},
   journal={Physical Review E},
   publisher={American Physical Society (APS)},
   author={Esposito, Massimiliano},
   year={2012},
   month=Apr }

@misc{mehl2012rolehiddenslowdegrees,
      title={Role of Hidden Slow Degrees of Freedom in the Fluctuation Theorem}, 
      author={Jakob Mehl and Boris Lander and Clemens Bechinger and Valentin Blickle and Udo Seifert},
      year={2012},
      eprint={1205.0238},
      archivePrefix={arXiv},
      primaryClass={cond-mat.soft},
      url={https://arxiv.org/abs/1205.0238}, 
}

@article{Zeraati_2012,
   title={Entropy production of nonequilibrium steady states with irreversible transitions},
   volume={2012},
   ISSN={1742-5468},
   url={http://dx.doi.org/10.1088/1742-5468/2012/12/L12001},
   DOI={10.1088/1742-5468/2012/12/l12001},
   number={12},
   journal={Journal of Statistical Mechanics: Theory and Experiment},
   publisher={IOP Publishing},
   author={Zeraati, Somayeh and Jafarpour, Farhad H and Hinrichsen, Haye},
   year={2012},
   month=Dec, pages={L12001} }

@article{Rahav_2014,
   title={An integral fluctuation theorem for systems with unidirectional transitions},
   volume={2014},
   ISSN={1742-5468},
   url={http://dx.doi.org/10.1088/1742-5468/2014/10/P10044},
   DOI={10.1088/1742-5468/2014/10/p10044},
   number={10},
   journal={Journal of Statistical Mechanics: Theory and Experiment},
   publisher={IOP Publishing},
   author={Rahav, Saar and Harbola, Upendra},
   year={2014},
   month=Oct, pages={P10044} }

@article{Busiello_2020,
   title={Entropy production in systems with unidirectional transitions},
   volume={2},
   ISSN={2643-1564},
   url={http://dx.doi.org/10.1103/PhysRevResearch.2.023011},
   DOI={10.1103/physrevresearch.2.023011},
   number={2},
   journal={Physical Review Research},
   publisher={American Physical Society (APS)},
   author={Busiello, D. M. and Gupta, D. and Maritan, A.},
   year={2020},
   month=Apr }

@misc{green2026quantummechanicsrareevents,
      title={The Quantum Mechanics of Rare Events: From Quantum Walks to Stochastic Inflation}, 
      author={Daniel Green and Kshitij Gupta and Akhil Premkumar},
      year={2026},
      eprint={2608.06319},
      archivePrefix={arXiv},
      primaryClass={hep-th},
      url={https://arxiv.org/abs/2608.06319}, 
}

@article{Ferri_Cort_s_2025,
   title={Conditional fluctuation theorems and entropy production for monitored quantum systems under imperfect detection},
   volume={7},
   ISSN={2643-1564},
   url={http://dx.doi.org/10.1103/PhysRevResearch.7.013077},
   DOI={10.1103/physrevresearch.7.013077},
   number={1},
   journal={Physical Review Research},
   publisher={American Physical Society (APS)},
   author={Ferri-Cortés, Mar and Almanza-Marrero, José A. and López, Rosa and Zambrini, Roberta and Manzano, Gonzalo},
   year={2025},
   month=Jan }

@article{Ciccarello_2022,
   title={Quantum collision models: Open system dynamics from repeated interactions},
   volume={954},
   ISSN={0370-1573},
   url={http://dx.doi.org/10.1016/j.physrep.2022.01.001},
   DOI={10.1016/j.physrep.2022.01.001},
   journal={Physics Reports},
   publisher={Elsevier BV},
   author={Ciccarello, Francesco and Lorenzo, Salvatore and Giovannetti, Vittorio and Palma, G. Massimo},
   year={2022},
   month=Apr, pages={1–70} }

@article{Strasberg_2017,
   title={Quantum and Information Thermodynamics: A Unifying Framework Based on Repeated Interactions},
   volume={7},
   ISSN={2160-3308},
   url={http://dx.doi.org/10.1103/PhysRevX.7.021003},
   DOI={10.1103/physrevx.7.021003},
   number={2},
   journal={Physical Review X},
   publisher={American Physical Society (APS)},
   author={Strasberg, Philipp and Schaller, Gernot and Brandes, Tobias and Esposito, Massimiliano},
   year={2017},
   month=Apr }

@article{Murashita_2014,
  author  = {Murashita, Yuto and Funo, Ken and Ueda, Masahito},
  title   = {Nonequilibrium equalities in absolutely irreversible processes},
  journal = {Physical Review E},
  volume  = {90},
  pages   = {042110},
  year    = {2014},
  doi     = {10.1103/PhysRevE.90.042110}
}

@article{Funo_2015,
  author  = {Funo, Ken and Murashita, Yuto and Ueda, Masahito},
  title   = {Quantum nonequilibrium equalities with absolute irreversibility},
  journal = {New Journal of Physics},
  volume  = {17},
  pages   = {075005},
  year    = {2015},
  doi     = {10.1088/1367-2630/17/7/075005}
}

@article{Manikandan_2019,
  author  = {Manikandan, Sreenath K. and Elouard, Cyril and Jordan, Andrew N.},
  title   = {Fluctuation theorems for continuous quantum measurements and absolute irreversibility},
  journal = {Physical Review A},
  volume  = {99},
  pages   = {022117},
  year    = {2019},
  doi     = {10.1103/PhysRevA.99.022117}
}

@article{Yuge_2020,
  author  = {Yuge, Tatsuro and Yamaguchi, Makoto},
  title   = {Fluctuation theorem in cavity quantum electrodynamics systems},
  journal = {Physical Review E},
  volume  = {101},
  pages   = {022113},
  year    = {2020},
  doi     = {10.1103/PhysRevE.101.022113}
}

@article{Kewming_2022,
   title={Entropy production and fluctuation theorems in a continuously monitored optical cavity at zero temperature},
   volume={6},
   ISSN={2521-327X},
   url={http://dx.doi.org/10.22331/q-2022-04-13-685},
   DOI={10.22331/q-2022-04-13-685},
   journal={Quantum},
   publisher={Verein zur Forderung des Open Access Publizierens in den Quantenwissenschaften},
   author={Kewming, Michael J. and Shrapnel, Sally},
   year={2022},
   month=Apr, pages={685} }

@article{Drummond1980,
  author    = {Drummond, P. D. and McNeil, K. J. and Walls, D. F.},
  title     = {Non-equilibrium Transitions in Sub/Second Harmonic Generation},
  journal   = {Optica Acta: International Journal of Optics},
  volume    = {27},
  number    = {3},
  pages     = {321--335},
  year      = {1980},
  publisher = {Taylor \& Francis},
  doi       = {10.1080/713820226}
}

@book{Walls_2008,
  author    = {Walls, D. F. and Milburn, G. J.},
  title     = {Quantum Optics},
  edition   = {2nd},
  publisher = {Springer},
  address   = {Berlin},
  year      = {2008}
}

\end{document}